# Theoretical relation between water flow rate in a vertical fracture and rock temperature in the surrounding massif

by Jean-Christophe Maréchal(✉)[1], Pierre Perrochet[2]

[1] Bureau de Recherches Géologiques et Minières, Service EAU, Unite RMD, 1029 avenue de Pinville, F-34000 Montpellier, France - BRGM-NGRI Indo-French Centre for Groundwater Research, NGRI, Uppal Road, 500 007 Hyderabad, India, Tel: + 91 40 715 80 90, Fax: + 91 40 717 15 64, marechal@satyam.net.in

[2] Centre d'Hydrogéologie, Université de Neuchâtel, 11 rue Emile Argand, CH-2007 Neuchâtel, Switzerland, Tel: + 41 32 718 25 77, Fax: + 41 32 718 26 03, pierre.perrochet@unine.ch

## *Abstract*

A steady-state analytical solution is given describing the temperature distribution in a homogeneous massif perturbed by cold water flow through a discrete vertical fracture. A relation is derived to express the flow rate in the fracture as a function of the temperature measured in the surrounding rock.

These mathematical results can be useful for tunnel drilling as it approaches a vertical cold-water bearing structure that induces a thermal anomaly in the surrounding massif. During the tunnel drilling, by monitoring this anomaly along the tunnel axis one can quantify the flow rate in the discontinuity ahead before intersecting the fracture.

The cases of the Simplon, Mont-Blanc and Gothard tunnels (Alps) are handled with this approach which shows very good agreement between observed temperatures and the theoretical trend. The flow rates before drilling of the tunnel predicted with the theoretical solution are similar in the Mont-Blanc and Simplon cases, as well as the flow rates observed during the drilling. However, the absence of information on hydraulic gradients (before and during drilling) and on fracture specific storage prevents direct predictions of discharge rates in the tunnel.

## *Introduction*

Hydrogeologists and geophysicists have observed and reported for a long time on the influence of groundwater circulation on rock temperature fields. The first studies in this domain showed the close relations between flow and heat transfers [1,2], and used these relations to determine the hydraulic properties of an aquifer [3], to prospect shallow aquifers [4] and determine groundwater circulation zones [5,6]. At a local scale, Bodvarsson [7] studied the effect of flow through a fracture on the temperature field in an impermeable embedding rock. At a regional scale, hydro-thermal effects have been analyzed by means of numerical modelling [8,9]. In the context of tunnel drilling in mountainous massifs, studies [10,11,12] have generally focused on temperature predictions along the tunnel axis without considering groundwater flow. More recent studies have been carried out to improve the prediction of water inflows into tunnels by means of indirect measurements, such as thermal anomalies, and modelling [13,14].

Thermal fields inside mountainous massifs not only depend on thermal rock characteristics and regional geothermal flux but also on topography and on possible water circulation in the massif. Cold water infiltrating at high elevations has cooling effects on the massif. These effects may be of different nature depending on how water flows through the massif. Regional dampening of geothermal gradients occurs in case of diffuse infiltration whereas local thermal anomalies are detected in case of flow concentrated in discrete features [15]. When properly monitored, the latter is particularly interesting since local thermal data may help characterise and assess the hydrological anomaly.

Such a situation was observed in the Mont-Blanc crystalline massif during the drilling of a road tunnel in 1960. A thermal anomaly was encountered while the drilling was approaching a sub-vertical, highly tectonised zone which produced dramatic water inflows in the tunnel (Fig. 1). Similar thermal phenomena have been observed in various Alpine tunnels [15].

Estimates of temperatures and water inflows in tunnels are some of the major engineering issues that must be solved before the methods of excavation, ventilation and drainage can be designed. The objective of this paper is to provide a first practical step in this direction by demonstrating that even a simplified hydro-thermal mathematical model can explain observed local thermal anomalies. The results presented here, however, correspond to flow rates and temperatures under natural conditions, i.e. without tunnel interaction.

## *Mathematical approach*

A theoretical relation between the discharge rate of water flowing in a vertical fracture and the temperature field in the surrounding rocks is presented below. Initially, hydro-thermal steady state is assumed.

Consider the modelled problem as shown in Fig. 2a. The rock massif is represented by a vertical semi-infinite section in the plane $(x, z)$, $0 \leq x \leq L$, $-\infty < z \leq H$, $z = H$ (m) corresponding to surface elevation. A vertical fracture with flow rate $Q$ (m$^2$ s$^{-1}$) is located at $x = x_f$ (m). Due to heat exchanges during the flow in the fracture, the water temperature in the fracture increases from surface temperature $T_0$ (C) to temperature $T_f$ (C) at depth $H$ ($z = 0$). Thus, $T_f$ is higher than $T_0$ but lower than the normal temperature $T_n$ that one would detect at this point without fracture flow.

The temperature distribution $T(x, z)$ in the domain is governed by the heat equation

$$\frac{\partial^2 T}{\partial x^2} + \frac{\partial^2 T}{\partial z^2} = 0 \qquad (1)$$

neglecting heat sources (radioactivity) in the rock, subject to appropriate boundary conditions.

At the surface a constant rock temperature equal to the average air temperature $T_0$ is assumed. Away from a fracture generating a thermal anomaly, the thermal flux is essentially vertical and, therefore, horizontal temperature gradients are neglected. In the model domain this is the case at x=0, where cooling effects are not detected and the temperature is normal. For the purpose of the analysis a virtual fracture with a very large, downward flow rate is considered at x=L, bringing the rock temperature uniformly down to surface temperature $T_0$. The boundary at x=L could also be seen as a vertical cliff at constant air temperature.

With the above assumptions, the boundary conditions are

$$T = T_0 \text{ at } z = H \qquad \frac{\partial T}{\partial x} = 0 \text{ at } x = 0 \qquad T = T_0 \text{ at } x = L \qquad (2)$$

and the solution of Eq.(1) is

$$T(x,z) = T_0 + (T_n - T_0)\frac{\cos\left(\frac{\pi x}{2L}\right)\sinh\left(\frac{\pi(H-z)}{2L}\right)}{\sinh\left(\frac{\pi H}{2L}\right)} \qquad (3)$$

where $T_n$ designates the temperature at $x = 0$ and $z = 0$. With no thermal anomaly, $T_n$ is the "normal" temperature that would be found along the axis $z = 0$. $T_n$ depends on the geothermal flux, the rock cover thickness, the rock thermal conductivity and the presence of diffuse water flows in the massif.

Along the axis $z = 0$, the longitudinal temperature profile is

$$T(x,0) = T_0 + (T_n - T_0)\cos(ax) \qquad (4)$$

where $a = \frac{\pi}{2L}$ characterises the strength of the anomaly.

Assuming now that the abnormal evolution of the temperature profile $T(x,0)$ over the distance $0<x< x_f$ is due to the presence of a real fracture at $x=x_f$, one may calculate the fracture flow rate at $x=x_f$, compatible with the observed anomaly between x=0 and $x=x_f$ where Eq. (4) is realist.

In this case, the temperature $T_f$ detected at $x_f$ (namely $T(x_f, 0)$) is the result of thermal equilibrium between advective heat transport in the fracture and cumulative conductive heat exchanges $F$ (W m$^{-1}$) along the fracture walls over the distance $H$. At $x = x_f$, the quantity $F$ is given by

$$F = 2\int_0^H \left(-k\frac{\partial T}{\partial x}\bigg|_{x=x_f}\right) dz = 2k(T_n - T_0)\sin(ax_f)\left(\frac{\cosh(aH)-1}{\sinh(aH)}\right) \qquad (5)$$

where $k$ is the thermal conductivity of the rock (W m$^{-1}$ K$^{-1}$).

By thermal continuity, and neglecting conductive transport within the fracture, the quantity $F$ is also

$$F = Q\rho_w(T_f - T_0) \qquad (6)$$

where $\rho_w$ (J m$^{-3}$ K$^{-1}$) is the volume thermal capacity of the water and $Q$ the fracture flow rate under natural conditions before the tunnel drilling.

Equating (5) and (6) yields

$$Q = \frac{2k}{\rho_w}\frac{(T_n - T_0)}{(T_f - T_0)}\sin(ax_f)\frac{(\cosh(aH)-1)}{\sinh(aH)} \qquad (7)$$

where, from Eq. (4)

$$ax_f = \arccos\left(\frac{T_f - T_0}{T_n - T_0}\right) \tag{8}$$

Given that

$$\sin\left(\arccos\left(U\right)\right) = \sqrt{1 - U^2} \tag{9}$$

the flow rate in Eq.(7) may finally be expressed as

$$Q = \frac{2k}{\rho_w}\sqrt{\left(\frac{T_n - T_0}{T_f - T_0}\right)^2 - 1}\;\frac{(\cosh(aH) - 1)}{\sinh(aH)} \tag{10}$$

If a tunnel is drilled along the $z = 0$ axis and the temperatures are regularly monitored by sounding at the front of the gallery, the parameters $T_n$, $T_f$ and $a$ in the above equation are known and the flow rate $Q$ in a potential fracture ahead can be evaluated.

In the case of a very deep tunnel for which $H > 2L$, Eq. (10) simplifies to

$$Q = \frac{2k}{\rho_w}\sqrt{\left(\frac{T_n - T_0}{T_f - T_0}\right)^2 - 1} \tag{11}$$

in which case the flow rate is not dependent on $a$, namely on $L$.

In Eq. (10) $\rho_w$ is constant ($4.2 \times 10^6$ J m$^{-3}$ K$^{-1}$), as well as $k$ for a given material. The variation ranges of the leading parameters $T_n$, $T_f$, $T_0$, $H$ and $L$ are given in Table 1. A set of typical values is also given with the corresponding discharge rate $Q_{ref}$ used to assess the sensitivity of the solution. Fig. 3 shows the evolution of the ratio $Q/Q_{ref}$ with the dimensionless variables $(T_n-T_o)/(T_f-T_o)$ and $H/L$. It is seen from this Figure that $Q$ is not influenced by a ratio $H/L > 2$ and that it scales quasi-linearly with $(T_n-T_o)/(T_f-T_o)$.

## *Application*

The analytical solution described above is applied to three field data sets gathered along the Simplon [16], the Mont-Blanc [17] and the Gothard [18] Alpine tunnels.

The Simplon railway tunnel was drilled at the beginning of the century through the Penninic nappes of the Southern Alps between Switzerland and Italy. After crossing a very hot zone ($T_{max} > 55°C$) with a thick rock cover, the temperatures of the gneissic low-permeability rocks stabilised around $T_n = 40°C$ (Km 13.048 from the North portal) before gradually decreasing when approaching the more permeable Teggiolo metasedimentary zone ($T_f = 18.2°C$ at Km 15.648). The analytical adjustment of the anomaly shown in Fig. 4 is done between x=0 and x=$x_f$ where Eq. (4) is realist, with $T_n = 40$ °C, $T_0 = 1°C$ assuming that the water temperature under glacier cover is near the melting point, and $L = 3600$ m ($a = 0.000436$ m$^{-1}$) in Eq. 4.

The Mont-Blanc road tunnel was drilled at the end of the fifties across the Mont-Blanc External Crystalline Massif between France and Italy. The temperature profile (Fig. 5) is typical with a negative thermal anomaly ($T_f = 11.5°C$) observed at Km 7.875 from the French portal in the middle of the massif ($T_n = 29.8°C$ at Km 6.270). This anomaly corresponds to a strongly fissured zone with large water inflows in the tunnel (almost 1000 l s$^{-1}$ for the total zone after [17]). The analytical adjustment of the anomaly shown in Figure 5 is done with $T_n = 29.8°C$, $T_0 = 1°C$ and $L = 2300$ m ($a = 0.000683$ m$^{-1}$).

The Gothard gallery, drilled between 1994 and 1997, was an exploration gallery for the Alptransit project in Switzerland. Its purpose was to investigate the hydrogeological and geomechanical behaviour of a metasedimentary rock syncline (strongly tectonised dolomites) named Piora Mulde in the middle of Leventina and Lucomagno Penninic gneisses. Temperatures reached $T_n$ = 31.5°C at Km 4.000 from the portal in the low-permeability gneiss and decreased to $T_f$ = 9.5°C in the neighbourhood of the metasedimentary rocks at the end of the gallery (5.595 Km from the portal). The analytical adjustment of the anomaly shown in Figure 6 is done with $T_n$ = 31.5°C, $T_0$ = 1°C and $L$ = 2020 m ($a$ = 0.000778 m$^{-1}$).

## *Discussion*

Despite the simplicity of the approach the above three anomaly adjustments are very satisfying. Parameters corresponding to these adjustments are given in Table 2 and flow rates are computed with $k$ = 3 W m$^{-1}$ K$^{-1}$ and $\rho_w$ = 4.2·10$^6$ J m$^{-3}$ K$^{-1}$ in Eq. (10).

Before drilling, the three sites were characterised by fracture discharge of similar magnitude of the order 10$^{-6}$ m$^3$ m$^{-1}$ s$^{-1}$ (0.9, 2.3 and 2.6 × 10$^{-6}$ m$^2$ s$^{-1}$ for the three sites respectively). Assuming a recharge rate of 150 mm yr$^{-1}$ (5 × 10$^{-9}$ m s$^{-1}$), these flow rates correspond to the concentration of a 200 m to 600 m-wide capture zone at the surface of the mountain.

The above flow rates depend on regional hydraulic gradients in natural conditions and are not necessarily correlated to the amounts of water that were locally drained by the tunnels during drilling. In fact, tunnel discharge is not only governed by fracture transmissivity but also by local drawdown as well as by the storage coefficient. Nevertheless, it is worth mentioning that, during drilling, discharge rates measured at early times were similar in the Mont-Blanc (1100 l s$^{-1}$) and Simplon (850 l s$^{-1}$) tunnels (no early-time data is available for the Gothard).

Although this consideration needs further experimental support, it could be inferred that, assuming a hypothetical regional hydraulic gradient typical of Alpine massifs, some correlation might exist, through fracture transmissivity, between flow rates derived from the present approach and early-time drainage in tunnels.

## *Conclusion*

An analytical solution describing steady-state hydro-thermal processes in the vicinity of a vertical fracture through a rock massif is given and applied to data sets from three Alpine tunnels. In the three cases, the computed flow rate corresponds to the concentration of the recharge to a 200 m to 600 m-wide capture zone subjected to typical mountain recharge rates.

Practically, during tunnel drilling, by fitting the observed temperature profiles on the proposed model it is possible to express the potential flow rate beyond the tunnel head as a function of the temperature at the tunnel head.

Furthermore, assumptions on typical regional hydraulic gradients provide a way to estimate fracture transmissivity and hence early-time potential discharge rates in the tunnel, applying specific analytical solutions of the problem of interaction between tunnel and groundwater. More experimental evidence is needed to confirm these preliminary findings.

## *Acknowledgments*

The authors thank Prof. L. Rybach and Prof. G. de Marsily for contributing to improve the present paper.

**TABLES**

| Parameters | $T_n - T_o$ (K) | $T_f - T_o$ (K) | $\frac{T_n - T_0}{T_f - T_0}$ (1) | $H$ (m) | $L$ (m) | $H/L$ (1) | $Q_{ref}$ (m² s⁻¹) |
|---|---|---|---|---|---|---|---|
| Variation | 10 - 50 | 0 - 20 | 1 - ∞ | 100 - 3000 | 100 - 3000 | 0.03 - 30 | - |
| Mean | 30 | 10 | 3 | 1000 | 1000 | 1 | $2.666 \times 10^{-6}$ |

*Table 1 : Parameters range frequently encountered in the field. $Q_{ref}$ is the flow rate computed with typical parameters (reference solution).*

| Site | $T_n$ (°C) | $T_o$ (°C) | $x_f$ (m) | $T_f$ (°C) | $H$ (m) | $a$ (m⁻¹) | $Q$ (m² s⁻¹) |
|---|---|---|---|---|---|---|---|
| Simplon | 40.0 | 1 | 2600 | 18.2 | 1500 | 0.000436 | $0.9 \times 10^{-6}$ |
| Mont-Blanc | 29.8 | 1 | 1605 | 11.5 | 2200 | 0.000683 | $2.3 \times 10^{-6}$ |
| Gothard | 31.5 | 1 | 1595 | 9.5 | 1500 | 0.000778 | $2.6 \times 10^{-6}$ |

*Table 2 : Parameters for the analytical adjustment of the three observed anomalies.*

**FIGURES**

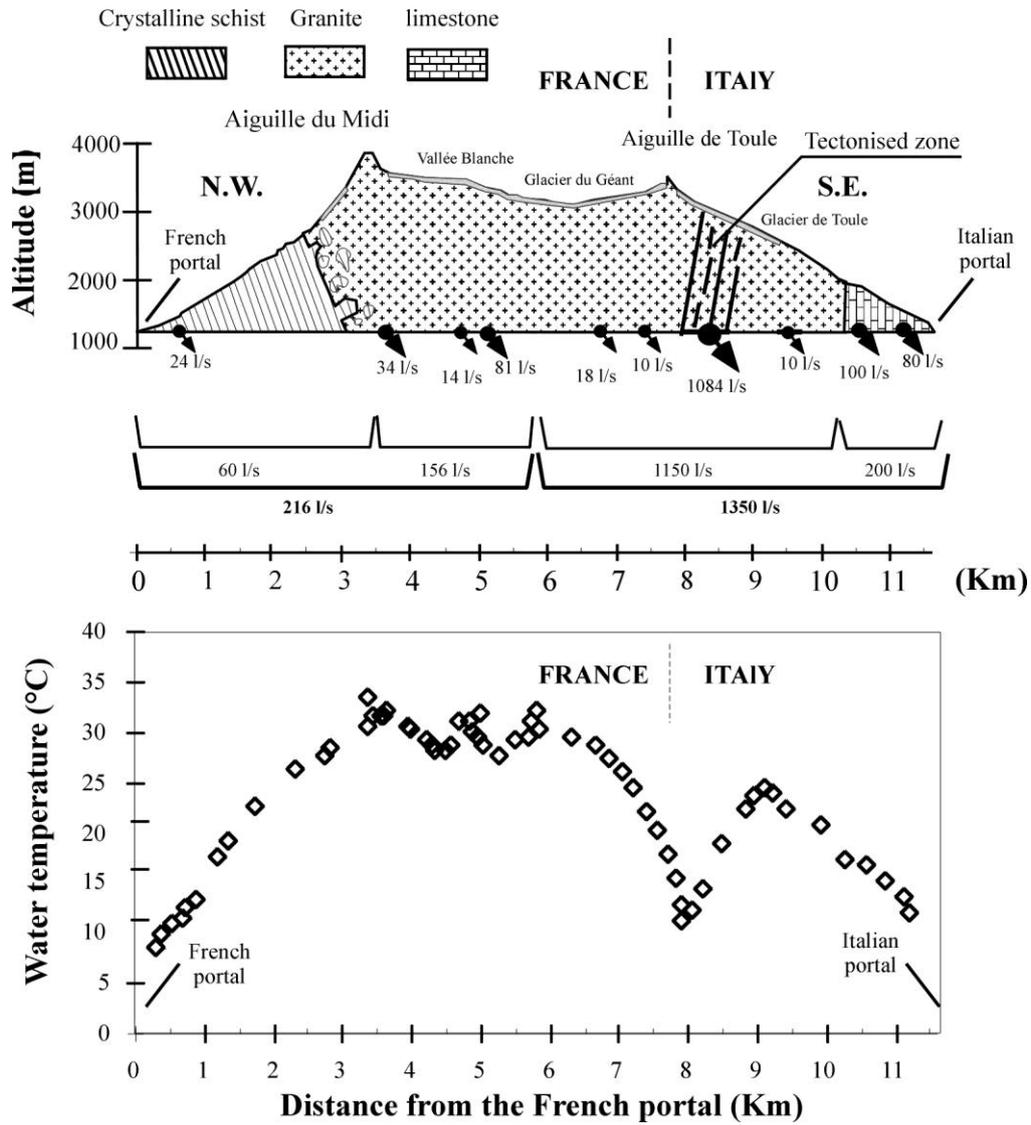

*Figure 1 : Schematic geology (top) and water temperatures (bottom) in the Mont-Blanc road tunnel.*

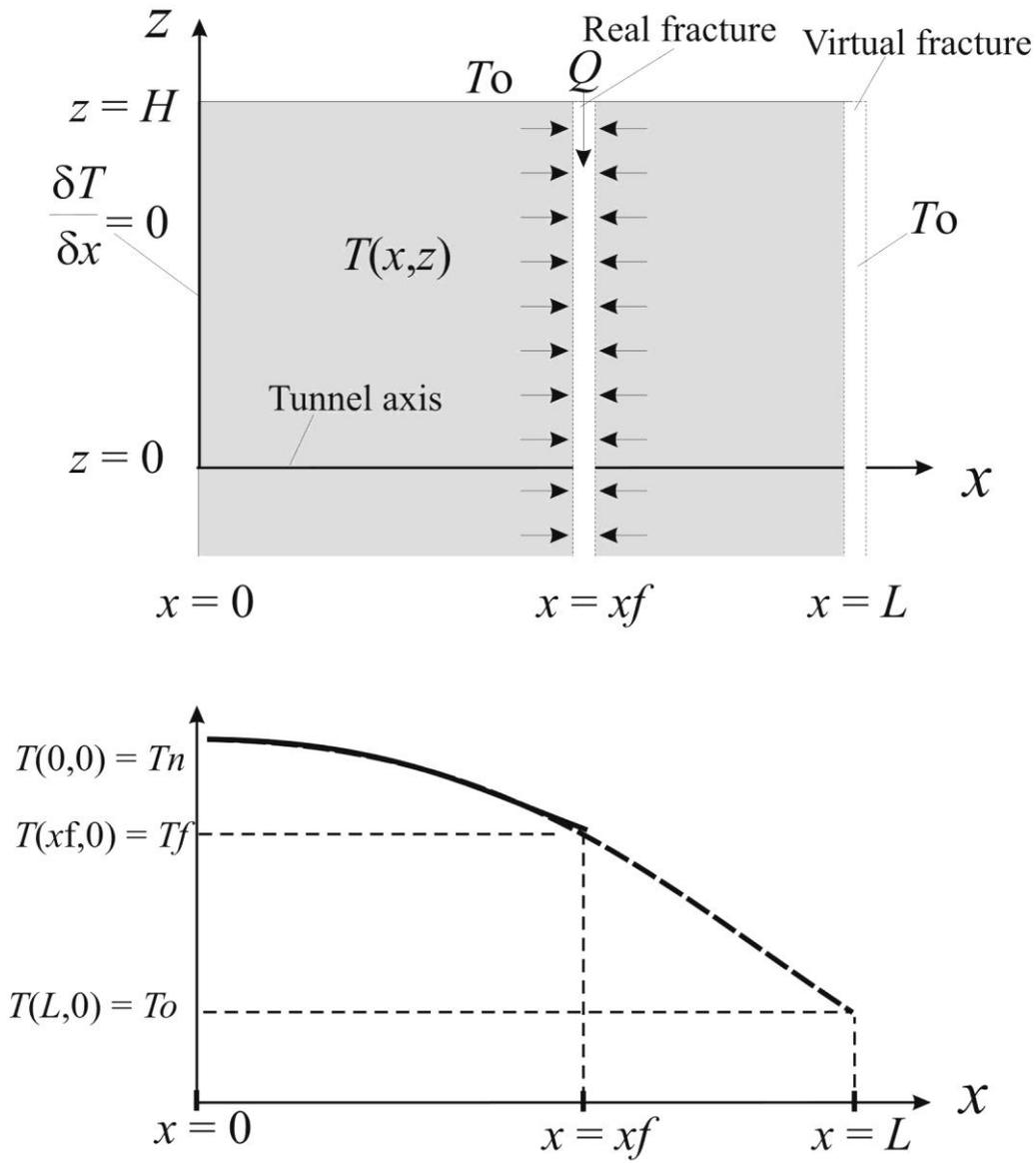

*Figure 2: sketch of the modelled problem (top) and temperature profile at z = 0 (bottom).*

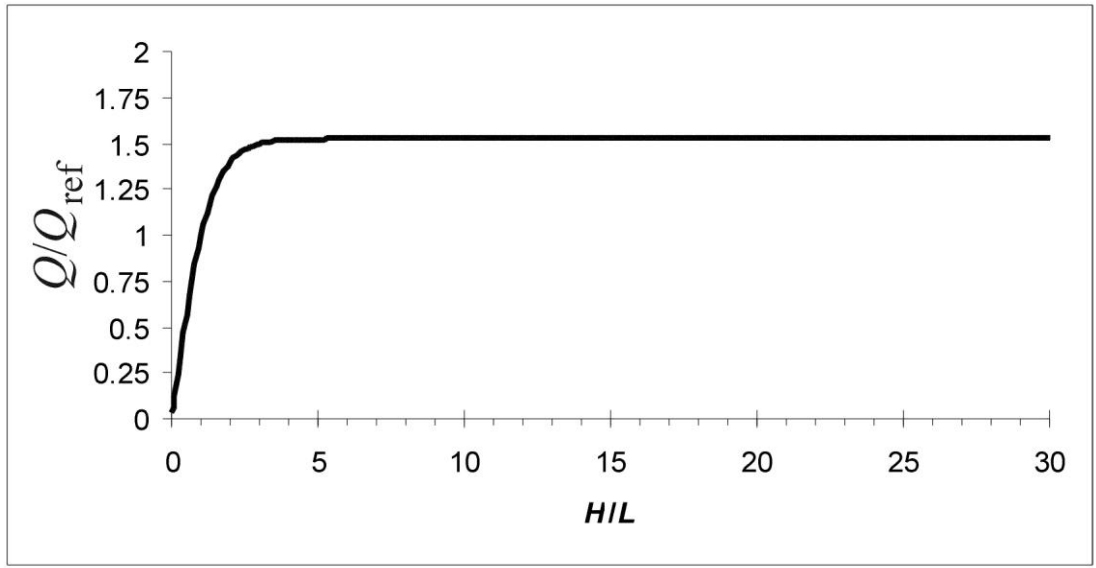

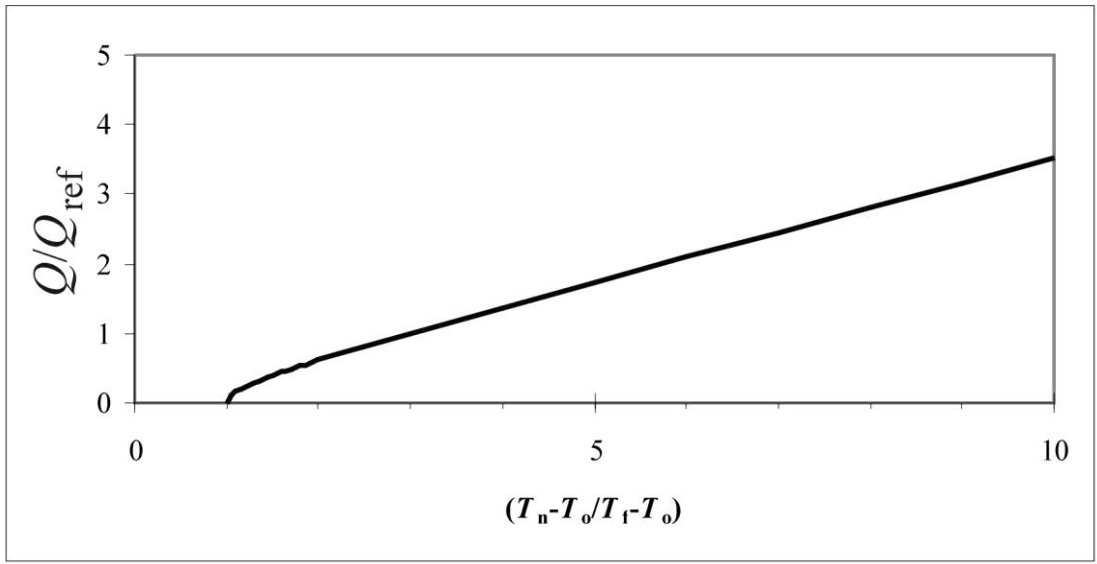

*Figure 3 : Computed flow rate versus H/L (top) and temperatures difference (bottom).*

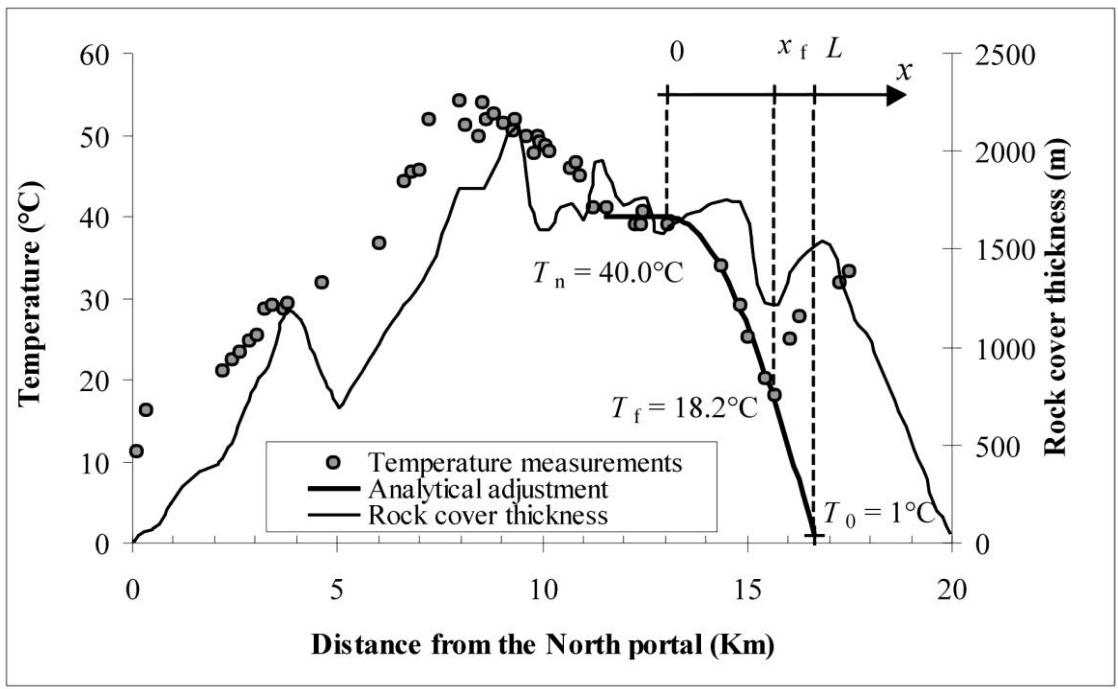

*Figure 4 : Water temperatures (measurements and analytical adjustment) and rock cover thickness in the railway Simplon tunnel.*

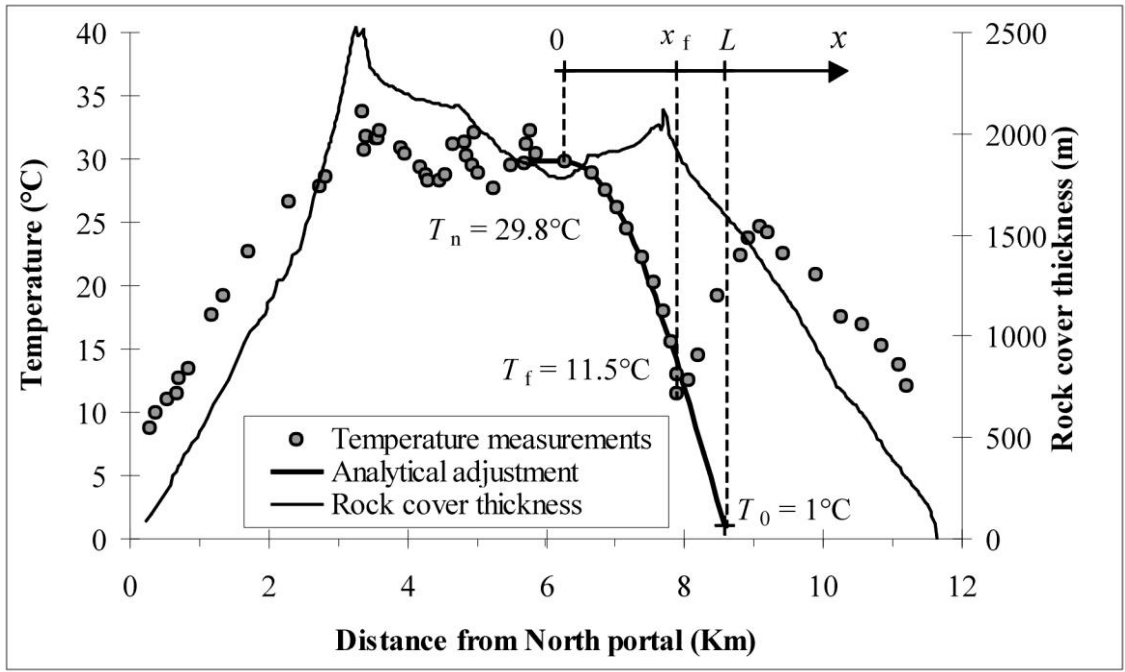

*Figure 5 : Water temperatures (measurements and analytical adjustment) and rock cover thickness in the Mont-Blanc road tunnel.*

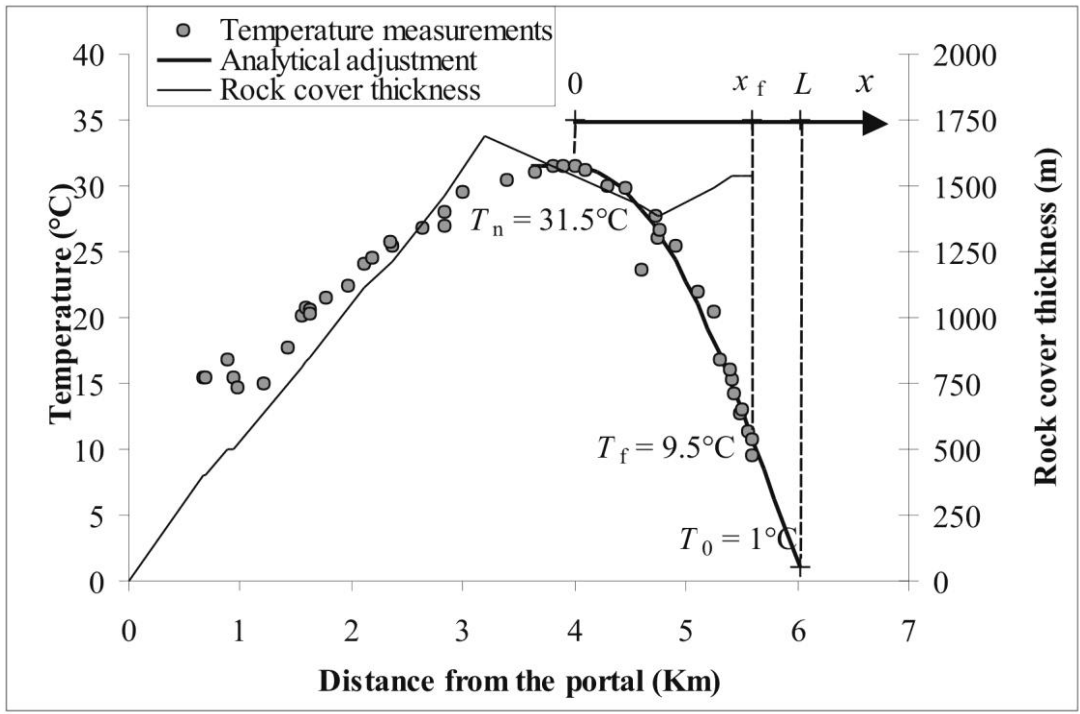

*Figure 6 : Water temperatures (measurements and analytical adjustment) and rock cover thickness in the Gothard gallery.*